# Unveiling a Novel Metal-to-Metal Transition in LuH$_2$: Critically Challenging Superconductivity Claims in Lutetium Hydrides

Dong Wang[1], Ningning Wang[2], Caoshun Zhang[1], Chunsheng Xia[1], Weicheng Guo[1], Xia Yin[1], Kejun Bu[1], Takeshi Nakagawa[1], Jianbo Zhang[1], Federico Gorelli[1], Philip Dalladay-Simpson[1], Thomas Meier[1], Xujie Lü[1], Liling Sun[1,2], Jinguang Cheng[2], Qiaoshi Zeng[1], Yang Ding[1*] and Ho-kwang Mao[1,3]

[1] Center for High-Pressure Science and Technology Advanced Research, Beijing, 100094, China
[2] Beijing National Laboratory for Condensed Matter Physics and Institute of Physics, Chinese Academy of Sciences, Beijing 100190, China
[3] Shanghai Key Laboratory of Material Frontiers Research in Extreme Environments (MFree), Shanghai Advanced Research in Physical Sciences (SHARPS), Shanghai 201203, China

*Correspondence：yang.ding@hpstar.ac.cn

Following the recent report by Dasenbrock-Gammon et al. (2023) of near-ambient superconductivity in nitrogen-doped lutetium trihydride (LuH$_{3-\delta}$N$_\varepsilon$), significant debate has emerged surrounding the composition and interpretation of the observed sharp resistance drop. Here, we meticulously revisit these claims through comprehensive characterization and investigations. We definitively identify the reported material as lutetium dihydride (LuH$_2$), resolving the ambiguity surrounding its composition. Under similar conditions (270-295 K and 1-2 GPa), we replicate the reported sharp decrease in electrical resistance with a 30% success rate, aligning with Dasenbrock-Gammon et al.'s observations. However, our extensive investigations reveal this phenomenon to be a novel, pressure-induced metal-to-metal transition intrinsic to LuH$_2$, distinct from superconductivity. Intriguingly, nitrogen doping exerts minimal impact on this transition. Our work not only elucidates the fundamental properties of LuH$_2$ and LuH$_3$ but also critically challenges the notion of superconductivity in these lutetium hydride systems. These findings pave the way for future research on lutetium hydride systems while emphasizing the crucial importance of rigorous verification in claims of ambient temperature superconductivity.

The 1911 discovery of superconductivity in mercury at 4.2 K, characterized by its vanishing electrical resistance, sparked a pivotal chapter in condensed matter physics[1]. This ignited an ongoing quest for materials exhibiting higher transition temperatures ($T_c$), a crucial factor for potential breakthroughs in critical technologies like energy transmission and transportation. The subsequent emergence of cuprates[2,3], boasting $T_c$ surpassing liquid nitrogen's boiling point (77.4 K), represented a significant leap forward. More recently, certain hydrides have also gained attention for displaying superconductivity above 200 K, albeit under extreme pressures exceeding 100 GPa[4-10].

In 2023, a particularly captivating yet controversial claim captured global attention. Dasenbrock-Gammon et al. reported the tantalizing observation of potential near-room-temperature superconductivity in a lutetium-hydrogen-nitrogen (Lu-H-N) compound at 294 K under modest pressure (1 GPa)[11]. However, the paper has since been retracted due to concerns about the reliability of the electrical resistance data and its analysis[12]. Consequently, the finding remains shrouded in uncertainty, with lingering questions surrounding the exact composition of the compound and the true nature of the observed decrease in electrical resistance. Further experimental verification and a deeper understanding of the underlying mechanisms are crucial to elucidating whether this preliminary study genuinely achieves the long-sought goal of ambient temperature superconductivity.

A significant controversy surrounds the composition of the proposed Lu-N-H compound. Dasenbrock-Gammon et al. propose that this phase is nitrogen-doped trihydride $LuH_{3-\delta}N$, attributing it to the observed blue-to-pink color transition and superconductivity. However, their claim contradicts established knowledge, which associates such color changes with the face-cubic-center (fcc) phase present in dihydride $LuH_2$[13-21]. While theoretical models suggest the possibility of nitrogen-doped fcc $LuH_3$ mimicking $LuH_2$[22-29], experimental evidence demonstrates that $LuH_3$ maintains its trigonal symmetry at pressures below 10 GPa[30]. Unfortunately, challenges in accurately quantifying hydrogen content hinder researchers' efforts to accurately determine the composition of the reported superconducting phase.

More importantly, the claimed near-ambient superconductivity of $LuH_{3-\delta}N_\delta$ at 294 K and 1 GPa remains unconfirmed by independent studies, casting doubt on its reality. Despite claims of a 35% replication success rate, subsequent investigations have failed to replicate these findings[13-18,31-34]. While some studies have observed a similar drop in electrical resistance, alternative explanations, including metal-insulator transitions[32] or percolative phenomena[34], have been proposed. These controversy highlights the crucial need for meticulous compositional analysis and rigorous validation before accepting any claim of superconductivity in this material.

To comprehensively investigate the controversial findings of Dasenbrock-Gammon et al., we meticulously replicated their synthesis protocol, exposing high-purity lutetium foil to a precisely calibrated 99:1 $H_2:N_2$ gas mixture under identical conditions of 2 GPa, 65 °C, and 24 hours[11]. Recognizing the inherent difficulty in differentiating between fcc $LuH_2$ and $LuH_3$ phases, we employed a separate strategy: synthesizing nitrogen-doped variants of both materials. We independently treated both $LuH_2$ and $LuH_3$ with nitrogen gas at 2 GPa for varying durations (1-5 days) at an elevated temperature of 200 °C to enhance the reaction kinetics. This approach enabled independent investigations of each material's properties, culminating in a comprehensive comparative analysis that clearly distinguished their unique characteristics.

Our comprehensive investigation, employing a combination of advanced analytical techniques including optical microscopy, X-ray diffraction, Raman spectroscopy, and electrical resistance measurements[35], yielded results significantly divergent from those of Dasenbrock-Gammon et al.[11]. We observed that lutetium metal remains largely unchanged under the prescribed $H_2:N_2$ gas mixture conditions of 2 GPa, 65 ℃, and 24 hours[11]. Importantly, nitrogen-doped $LuH_3$ retained its insulating, grey, and trigonal characteristics even under pressures exceeding 10 GPa. These findings offer compelling evidence that both the fcc metallic phase and the blue-to-pink color changes previously attributed to $LuH_{3-\delta}N_\varepsilon$ are inherent to $LuH_2$.

Crucially, electrical resistance measurements on both nitrogen-doped and undoped $LuH_2$ samples revealed a significant drop at 270-295 K and 1-2 GPa, with a success rate of 30%, consistent with observations by Dasenbrock-Gammon et al. However, our further experiments indicate that this reduction corresponds to a metal-to-metal phase transition rather than superconductivity. Consequently, we assert that the observed decrease in electrical resistance is more accurately attributed to a metal-to-metal transition in $LuH_2$, rather than indicative of superconductivity in $LuH_{3-\delta}N_\varepsilon$. Further details are provided in the subsequent sections.

**Reproducibility of the Published Protocol**

Figures 1a and 1b illustrate characterization results obtained by meticulously adhering to the original synthesis protocol outlined by Dasenbrock-Gammon et al. Unexpectedly, lutetium metal chips approximately seven microns thick retain their original color even when exposed to a 99:1 $H_2:N_2$ gas mixture at pressures exceeding 3.0 GPa.

X-ray diffraction (XRD) analyses, presented in Figure 1d, reveal the dominant presence of unreacted lutetium metal, contradicting the anticipated formation of a blue, nitrogen-doped lutetium hydride phase. Additionally, high-pressure electrical resistance profiles, shown in Figures 1e and 1f, exhibit a typical metallic behavior characterized by linear temperature dependence above 50 K, thus challenging prior claims of superconducting phase transitions in nitrogen-doped lutetium hydrides.

Recent studies suggest that extending the reaction time[32] or employing thinner lutetium metal foils[33] promotes the formation of lutetium hydrides, highlighting the slow kinetics of the reaction at these pressure-temperature conditions.

**Color and Structure of Nitrogen-Doped $LuH_3$**

$LuH_3$ exhibits trigonal symmetry under ambient conditions and transitions to a cubic phase above 10 GPa[30]. While Dasenbrock-Gammon et al. proposed that trace nitrogen might stabilize this cubic metallic $LuH_3$ phase below 2 GPa and 65 ℃, this proposition has not yet been experimentally confirmed.

To investigate this, we conducted experiments exposing $LuH_3$ to $N_2$ gas for 24 hours at 2.0 GPa and 200.0 ℃ (Fig. 2a). Notably, the temperature was elevated by 135 ℃ compared to the 65 ℃ used by Dasenbrock-Gammon et al. to further enhance reaction kinetics. Even at elevated pressures exceeding 30.2 GPa, the intrinsic dark grey hue of LuH$_3$ remains unchanged (Fig. 2b and S1). XRD and Raman spectroscopic analyses (Fig. 2c and 2d) corroborate that the crystalline lattice largely retains its initial trigonal symmetry after heat treatment at 2 GPa. Furthermore, the electrical resistance of nitrogen-doped LuH$_3$ remains within a consistent range of $10^5$ - $10^7$ Ohm ($\Omega$) up to pressures approaching 9.3 GPa (Fig. 2e and 2f), reinforcing its intrinsic insulating behavior. Increasing the reaction time between LuH$_3$ and N$_2$ gas to 5 days did not lead to any noticeable changes.

These data collectively cast doubt on the hypothesis that trace nitrogen quantities could stabilize a metallic fcc LuH$_{3-\delta}$N$_\epsilon$ phase, which was purported to undergo a color change from blue to pink at pressures exceeding 0.3 GPa[11].

**Composition of Reported Compound: LuH$_3$ or LuH$_2$**

Figure 3 explores the influence of nitrogen on LuH$_2$ under consistent experimental conditions. Notably, the pressurized nitrogen-doped LuH$_2$ exhibits a striking color change from blue to pink or red (Fig. 3a and S2). Single-crystal XRD and Raman spectroscopy (Fig. 3b and 3c) confirm the stabilization of this phase into a fcc structure. This fcc structure exhibits a lattice parameter of 5.0378 Å, compared to the pristine LuH$_2$ sample's 5.0235 Å. This lattice expansion suggests the incorporation of nitrogen atoms into the crystal lattice. In addition, electrical resistance measurements (Fig. 3d) reveal a conventional metallic state in nitrogen-doped LuH$_2$ at 0.4 GPa, without significant variations. This pressure is considerably lower than the 1 GPa reported for superconductivity[11]. These results indicate that the structure and properties of nitrogen-doped LuH$_2$ closely resemble those of pristine LuH$_2$[13,36].

Our comprehensive data demonstrate that nitrogen-doped LuH$_2$ adopts an fcc structure, undergoes pressure-induced color changes, and exhibits metallic behavior at low pressures. In stark contrast, nitrogen-doped LuH$_3$ retains its native trigonal lattice and lacks corresponding color changes and metallic behavior. Comparative analysis (Fig. 4a-f) reveals a striking consistency between the structure and vibrational properties of the Lu-H-N compound reported by Dasenbrock-Gammon et al. and those of fcc-structured LuH$_2$. This compelling evidence suggests that the reported Lu-N-H compound is LuH$_2$ rather than LuH$_3$. Furthermore, our investigation reveals minimal influence of nitrogen doping on both LuH$_2$ and LuH$_3$, further supporting this conclusion.

**Nature of the reported Sharp Drop in Electrical Resistance**

Across various LuH$_2$ samples, including nitrogen-doped, undoped, polycrystalline, and single-crystal forms, Figures 5a-c demonstrate a sharp decrease in electrical resistance with decreasing temperature. This occurs within a temperature range of 270-295 K and pressures between 1-2 GPa, without altering the crystal structures. Importantly, the transition temperature, pressure, and rate of resistance reduction align with the 294 K and 1 GPa results reported by Dasenbrock-Gammon et al. (Fig. 5d). However, zero resistance is not achieved.

Employing the background subtraction approach of Dasenbrock-Gammon et al., our data in Figures 5e-g seemingly exhibit zero resistance, hinting at potential superconductivity in line with their study. Interestingly, around 30% of our $LuH_2$ samples exhibit this resistance decrease, similar to the 35% previously reported by Dasenbrock-Gammon et al. Although successful replication has been limited, the consistency of pressure and temperature conditions associated with the resistance drop across independent experiments in two different laboratories offers compelling evidence that this phenomenon reflects an electronic phase transition.

In our exploration of potential superconductivity in nitrogen-doped $LuH_2$, we employed Andreev reflection[37,38] (Figures 5i-j) as a diagnostic method, in line with established protocols[11]. This quantum mechanism enables electrons from a normal metal to pair as Cooper pairs and cross into a superconductor, typically signaling the presence of superconducting states. While Andreev reflection is prominently observable in established superconductors such as $MgB_2$, our experiments with nitrogen-doped $LuH_2$ did not reveal any distinct Andreev reflection signatures.

It is crucial to understand that the absence of clear Andreev reflection indicators does not definitively preclude the existence of superconductivity. The reliability of Andreev reflection for diagnosing superconductivity is heavily influenced by the characteristics of the tunneling barrier at the superconductor-normal metal interface. Although the lack of Andreev reflection alone does not conclusively dismiss the possibility of superconductivity, our comprehensive observations, including the absence of zero resistance, strongly imply that the reduction in electrical resistance observed in nitrogen-doped $LuH_2$ is not characteristic of superconducting behavior.

Previous studies have reported that mechanical grinding of polycrystalline $LuH_2$ at ambient conditions can lead to metal-to-insulator transitions[32] and percolation[34]—the abrupt transition from a non-conducting to a conducting state upon reaching a critical density of conductive elements. This phenomenon is typically associated with inhomogeneous systems like polycrystalline or granular materials. In contrast, our data reveal a distinct drop in electrical resistance in single-crystal $LuH_2$ samples within 270-295 K and 1-2 GPa. This signifies a metal-to-metal electronic transition rather than a transition to an insulator.

It is well-established that achieving true stoichiometric fcc $LuH_2$ is challenging. Hydrogen vacancies often form during synthesis, resulting in non-stoichiometric $LuH_{2+x}$ compounds. These vacancies significantly impact resistivity and optical properties while leaving the overall crystal structure largely unchanged. This behavior is characteristic of rare-earth metal hydrides and has been extensively studied in materials like 'switchable mirrors'[39-45] that switch between transparent insulating and reflective metallic states.

Therefore, the metal-to-metal electronic transition observed in our study likely arises from alterations in hydrogen-lutetium interactions influenced by factors such as pressure, strain/stress, temperature, and the spatial distribution of hydrogen vacancies. These external variables could explain the 30% -35% replication rate observed. Additionally, our investigation revealed that nitrogen doping has a negligible impact on the observed metal-to-metal transition. Encapsulating a single-crystal of $LuH_2$ in a nitrogen atmosphere within a diamond anvil cell at 1-2 GPa for five days resulted in no observable changes in electrical resistance.

In summary, our research refutes claims of near-ambient-condition superconductivity in nitrogen-doped $LuH_{3-\delta}N_\varepsilon$. Data shows features attributed to $LuH_{3-\delta}N_\varepsilon$ are intrinsic to $LuH_2$, including the metallic fcc phase and pressure-induced color changes. We successfully replicated the resistance drop in LuH$_2$ at comparable conditions with similar success rate to Dasenbrock-Gammon et al. However, our extensive investigations reveal this phenomenon to be a novel, pressure-induced metal-to-metal transition intrinsic to LuH$_2$, distinct from superconductivity. Additionally, nitrogen minimally impacts this transition.

**Methods:**

**Synthesis of Lu-H-N samples.** Considering that commercially available Lu metal foils often contain $LuH_2$ and $Lu_2O_3$, we used X-ray diffraction (XRD) to rigorously select the purest foils based on their diffraction patterns. All samples were handled within a glovebox and prepared with gas in a pre-purged, hydrogen-rich environment, following the procedures detailed in the original report. The conditions within the glovebox were controlled to ensure that $O_2$ and $H_2O$ levels were consistently below 0.5 ppm. Subsequently, we heated the sample overnight in an oven at 65 °C, and after 24 hours, we opened a DAC to retrieve the sample.

We also used commercially available cubic $LuH_2$ single crystals as precursors to synthesize nitrogen-doped $LuH_2$. Before loading with nitrogen gas, we performed a single crystal XRD measurement to assess its quality. Several high-purity $LuH_2$ crystals were placed into a diamond anvil cell (DAC), loaded with nitrogen gas, and subjected to pressure increase up to 2.0 GPa. The DAC was then kept in an oven at 200 °C for 24 hours. Once the heat treatment concluded, single crystals were removed for further measurements. The synthesis process of nitrogen-doped $LuH_3$ was identical to that of $LuH_2$, using commercially available $LuH_3$ powders as precursors, loaded into a diamond anvil cell, pressurized to 2.0 GPa, and kept in an oven at 200 °C for 24 hours.

**Single crystal and powder XRD measurements.** Powder and single crystal XRD measurements were conducted on a Bruker D8 Venture diffractometer utilizing Mo Kα radiation. We collected the powder XRD rings of Lu foil and $LuH_3$ powders, both pre- and post-heating, using a charge-coupled device detector. These data were then integrated into XRD patterns with the assistance of APEX3 software. For the high-pressure powder XRD measurements of $LuH_3$, both pre- and post-heating, samples were prepared in SC-type DACs with rhenium gaskets. We calibrated the distance and tilt of the detector using $CeO_2$ powder. For $LuH_2$ and nitrogen-doped $LuH_2$, appropriate single crystals were selected for data collection, which was conducted at room temperature. The crystal structures of $LuH_2$, both pre- and post-heating, were solved and refined using the APEX3 software.

**Raman Spectroscopy measurements.** Raman spectra of Lu, $LuH_2$, $LuH_3$, and corresponding reactants were collected using a S&I MonoVista CRS+ Raman system. The 532 nm laser was utilized for the excitation, the laser power ranges from 1.148 mW to 4.572 mW, 300 grating/mm or 2400 grating/mm was employed during the measurements. Raman system was calibrated with the single crystal silicon (520 $cm^{-1}$) before measurements.

**High pressure electrical transport and Andreev reflections measurements.** The resistance of the as-synthesized samples under high pressure was measured using the Van der Pauw method in BeCu alloy symmetric DACs with a culet size of 300 $\mu$m. For each electrical transport measurement, a pre-pressed and drilled Re gasket insulated by $c$BN/epoxy mixtures was employed, creating a hole with a diameter of 100 $\mu$m. Nitrogen gas and NaCl powder were used as pressure transmitting medium. Four Pt strips served as conductive wires, with ruby functioning as the pressure marker. Temperature-dependent resistance measurements were conducted on an electrical transport system, equipped with a Keithley 6221 current source, a 2182A nanovoltmeter, and a 7001 switch device. For high-pressure Andreev reflection measurements, sharp Pt tips were cut to make contact with the sample. Prior to measuring single-crystal $LuH_2$ in nitrogen, the Andreev reflection experiment on $MgB_2$ was conducted below 60 K using our electrical transport system.

**Acknowledgement**

We are grateful for the help from F. Liu and L. Yang for the help on the X-ray diffraction experiments. Y. Ding is also grateful for the support from the National Key Research and Development Program of China (Grant Nos. 2022YFA1402301and 2018YFA0305703) and the National Natural Science Foundation of China (Grant Nos. U2230401). J.G. Cheng is supported by the National Key R&D Program of China (Grant No. 2021YFA1400200), the National Natural Science Foundation of China (Grant Nos. 12025408, 11921004) and the Strategic Priority Research Program of CAS (XDB33000000).

**Author contributions**

Y.D conceived and designed the experiments. D.W performed electrical resistance, Raman, and X-ray experiments. N.W, C.Z, C.X, W.G, X.Y, K.B, T.N, J. Z, F.G, P. D, T.M, X.L, L.S, J. C, all participated and assisted the experiments. D.W and Y.D analyzed the data. Y.D, D.W, H.M, and Q.Z discussed and interpreted the results. Y.D, D.W, and H.M wrote the paper.

**Competing interest.** The authors have no conflicts to disclose.

**Data Availability:** The data presented in this work are available upon reasonable request.

Figures

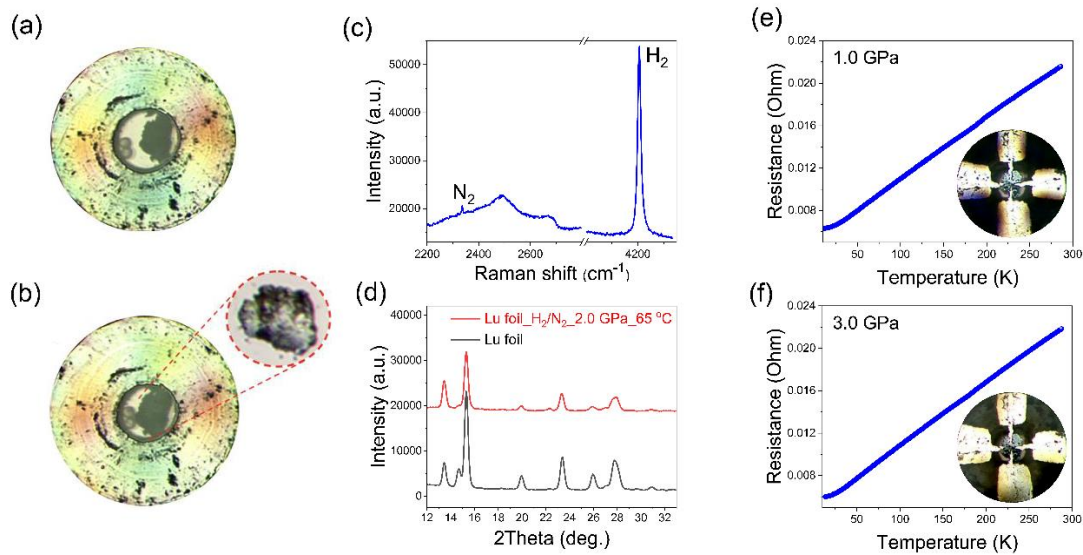

Figure 1. Synthesis and characterization of Lu metal under a $H_2:N_2$ (99:1) gas mixture following established protocol. (a) Pre-heating phase of the lutetium sample at 2.0 GPa with a 99% $H_2$ and 1% $N_2$ gas mixture at 1.0 GPa; (b) Post-heating phase for the same sample, maintained at 65 ℃ and 2.0 GPa for 24 hours, conspicuously devoid of the blue hue reported by Dasenbrock-Gammon et al.; (c) Raman spectrum of the gas mixture in contact with the lutetium sample; (d) X-ray diffraction (XRD) patterns for pure lutetium metal compared with the sample post 24-hour heating at 65 ℃ and 2.0 GPa; (e) Resistance variations with temperature for the sample post-heating at 1.0 GPa; (f) Resistance variations with temperature for the sample post-heating at 3.0 GPa.

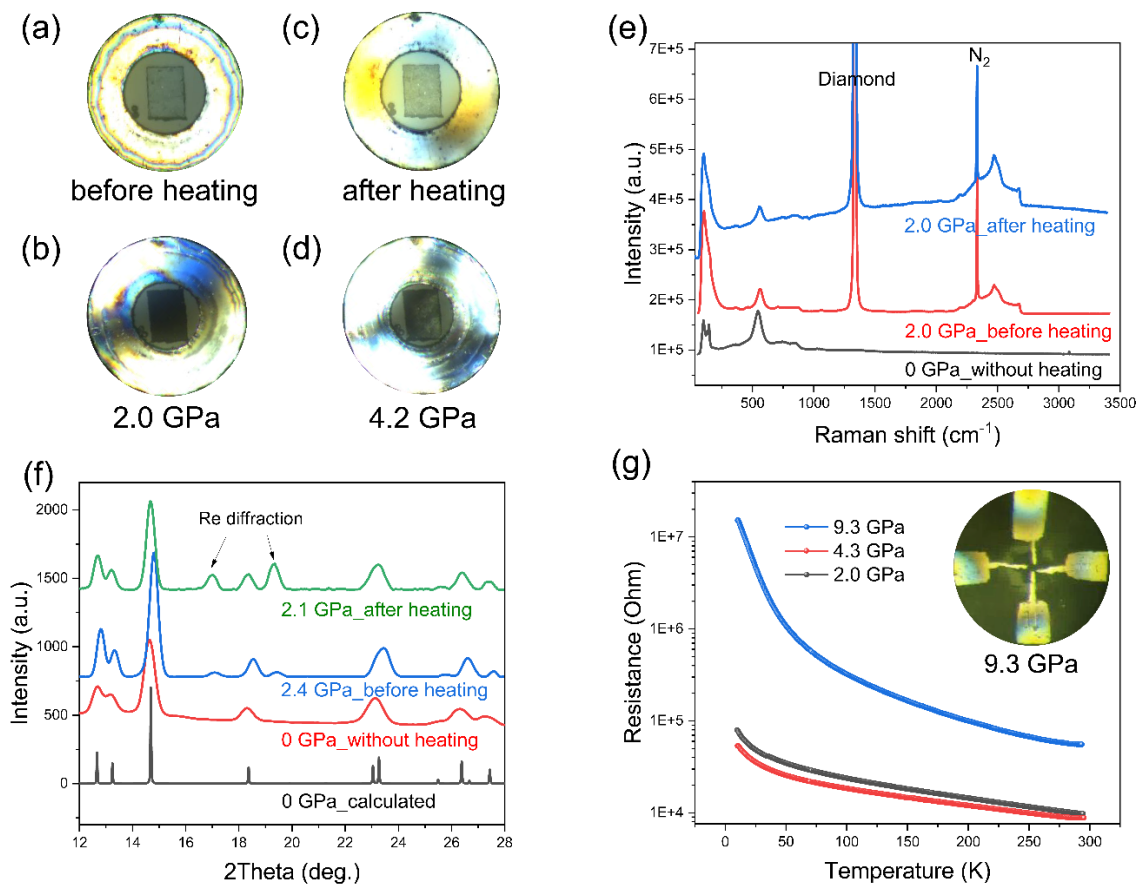

Figure 2. Characterization of nitrogen-doped LuH$_3$ synthesized at 200 °C and 2.0 GPa for 24 Hours. (a-d) Sequential images document the sample's color evolution during synthesis, with no observed blue-to-pink color changes. (e) Comparative Raman spectra of LuH$_3$ before and after pressurized heating; undoped LuH$_3$ spectra included for reference. (f) XRD patterns of LuH$_3$ pre- and post-heating, complemented by the un-doped LuH$_3$ pattern at 0 GPa and a simulated LuH$_3$ pattern. (g) Resistance as a function of temperature for nitrogen-doped LuH$_3$ at pressures from 2.0 GPa to 9.3 GPa. The inset illustrates the four-probe measurement.

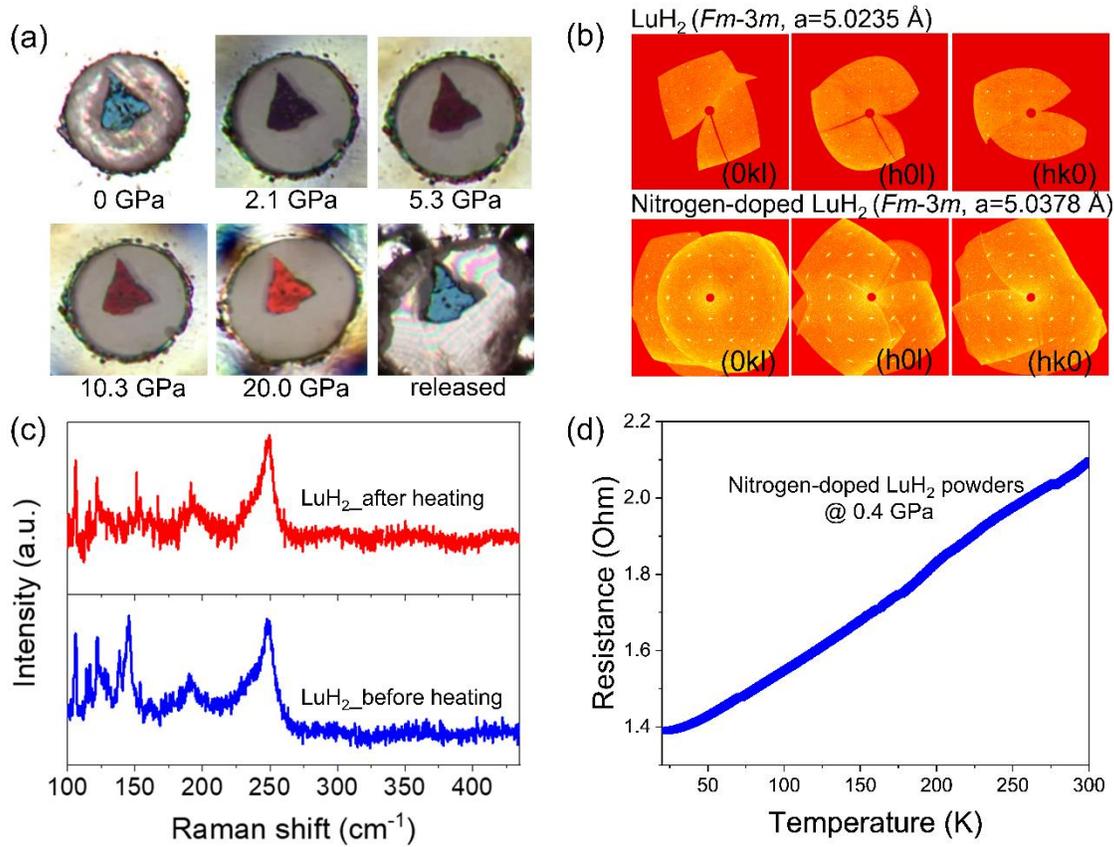

Figure 3. Characterization of nitrogen-doped $LuH_2$ produced from single-crystal $LuH_2$ treated in $N_2$ gas at 200 ℃ and 2.0 GPa. (a) Sequential photographs capture the sample's color evolution as pressure ranges from 0 to 20.0 GPa. (b) XRD patterns of both un-doped and nitrogen-doped $LuH_2$ single-crystals. (c) Raman spectra of $LuH_2$ pre- and post-heat treatment. (d) Electrical transport behavior of nitrogen-doped $LuH_2$ powders under 0.4 GPa.

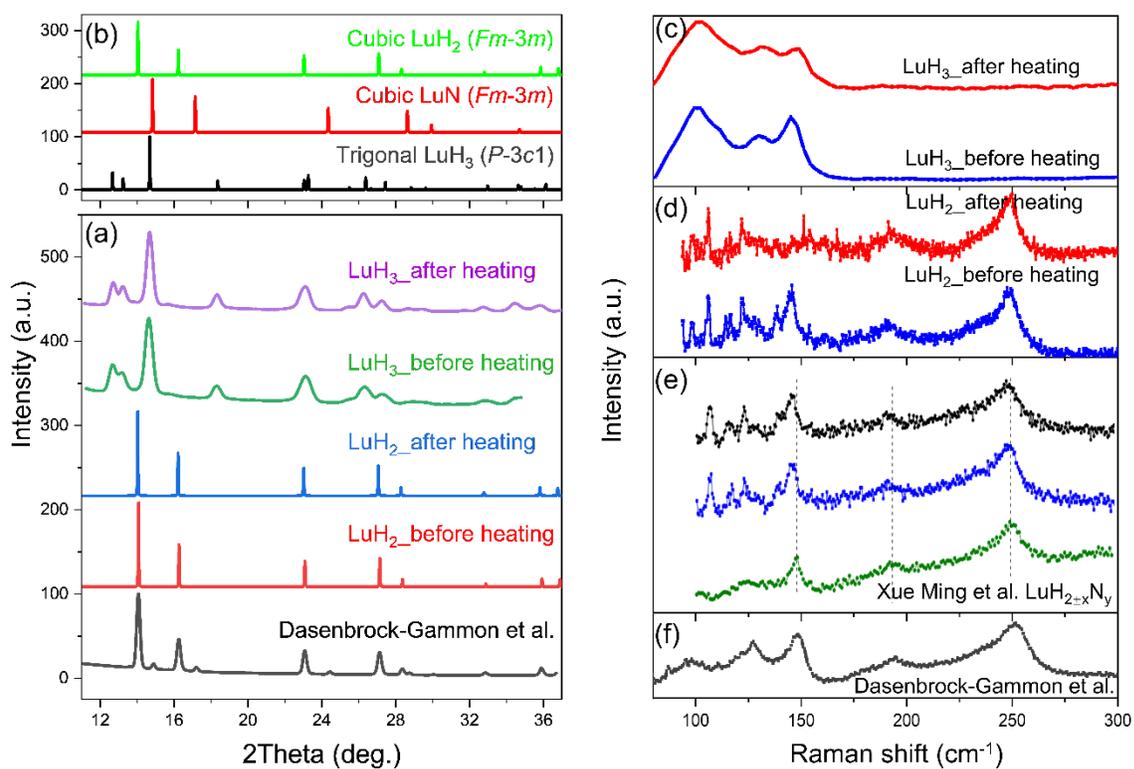

Figure 4. Detailed comparison of structure and properties of $LuH_2$ and $LuH_3$. (a) The XRD patterns of $LuH_2$ and $LuH_3$ before and after heating, along with the diffraction pattern of the Lu-N-H compound; (b) The simulated XRD pattern of $LuH_2$ ($Fm\bar{3}m$), LuN ($Fm\bar{3}m$), and $LuH_3$ ($P\bar{3}c1$); (c) Raman spectra of $LuH_3$ before and after heating; (d) Raman spectrum of $LuH_2$ before and after heating; (e) Raman spectra of as-synthesized $LuH_{2\pm x}N_y$ compounds from Xue Ming et al.[18]; (f) Raman spectrum of the Lu-H-N compound[11].

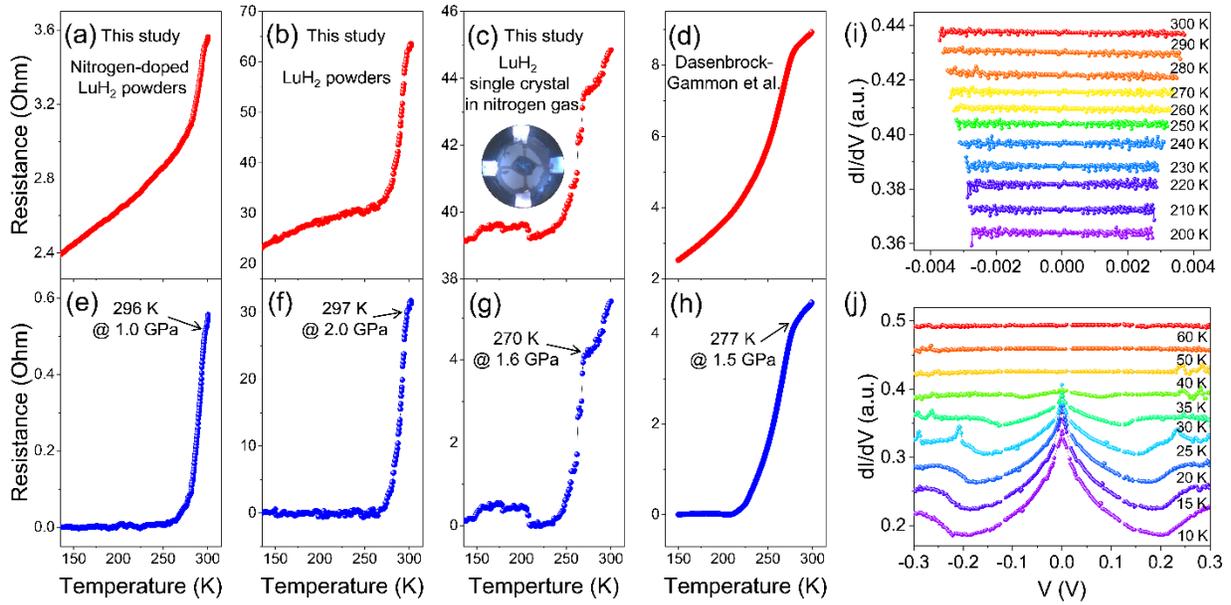

Figure 5. Comparative electrical resistance measurements of nitrogen-doped and pristine $LuH_2$ in various forms and under pressures in range of 1-2 GPa. (a) Electrical resistance of nitrogen-doped, polycrystalline $LuH_2$, illustrating a drop at 296 K and 1.0 GPa; (b) Electrical resistance of pristine, polycrystalline $LuH_2$, showing a drop at 297 K and 2.0 GPa; (c) Electrical resistance of single-crystal $LuH_2$ in a nitrogen gas, indicating a drop at 270 K and 1.6 GPa. Inset: Optical photograph of the four-probe electrical resistance setup in a diamond anvil cell; (d) Data from Dasenbrock-Gammon et al., exhibiting a resistance drop at 296 K and 1.0 GPa; (e)-(h) Resistance profiles after background subtraction, following the methodology of Dasenbrock-Gammon et al.; (i) Andreev reflection measurements from single-crystal $LuH_2$ under 1.6 GPa in nitrogen; (j) Andreev reflection measurements from standard $MgB_2$ for comparison, conducted under identical experimental conditions.

Supplementary materials

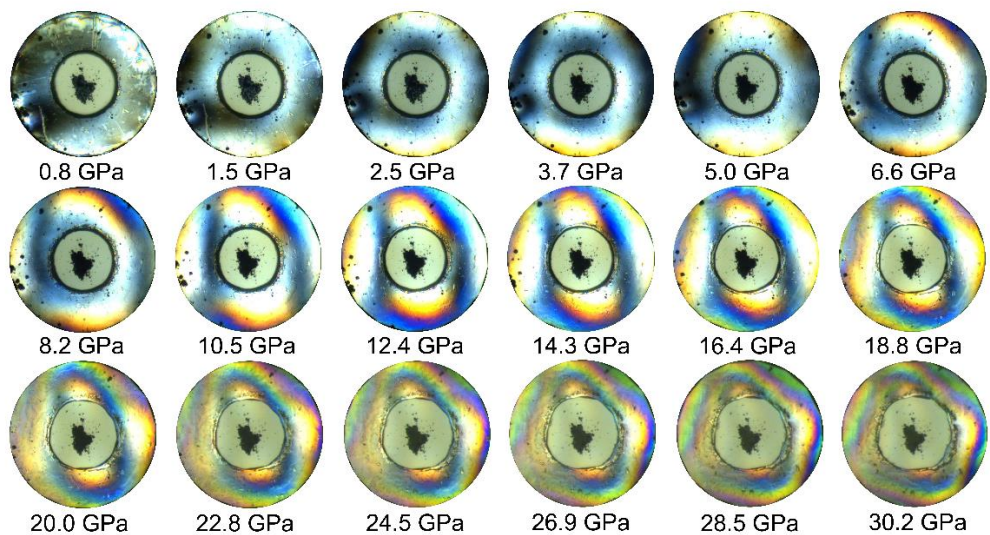

Figure S1. Pressure-dependent color changes of nitrogen-doped LuH$_3$.

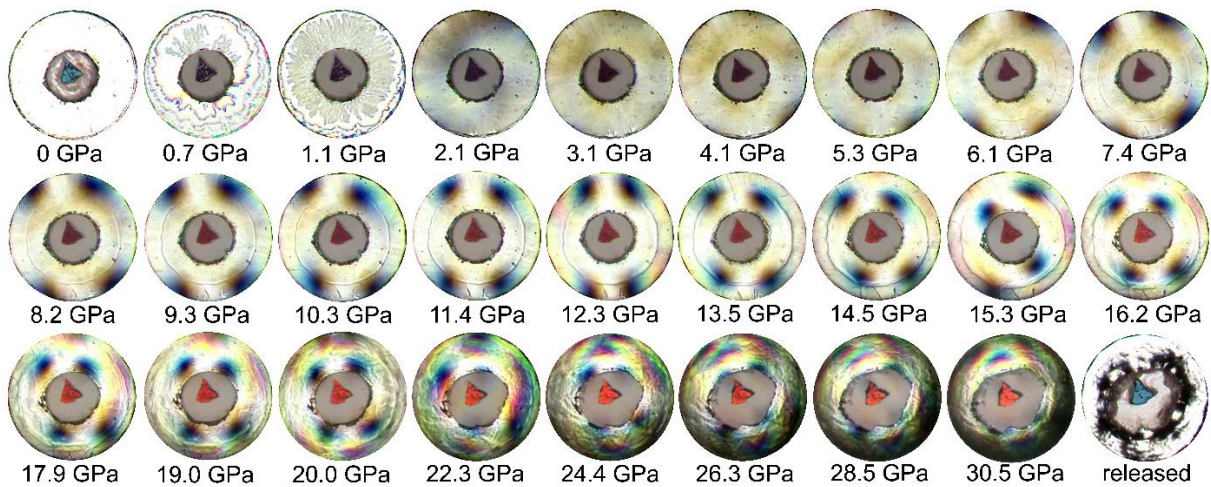

Figure S2. Pressure-dependent color changes of nitrogen-doped LuH$_2$.